\documentstyle{aipproc}
\begin{document}
\begin{flushright}
{\small
SLAC--PUB--7965\\
October 1998\\}
\end{flushright}
\vspace{8mm}
\title{
Dynamic Aperture Studies for SPEAR~3~\thanks
{Work supported by the Department of Energy Contract
DE-AC03-76SF00515 and the Office of Basic Energy Sciences, Division
of Chemical Sciences.}
}
\author{
Y.~Nosochkov and J.~Corbett
}
\address{
Stanford Linear Accelerator Center, Stanford University, Stanford, CA 94309
}
\maketitle
\begin{abstract}
The Stanford Synchrotron Radiation Laboratory is investigating an 
accelerator upgrade project that would replace the present 130~nm$\cdot$rad
FODO lattice with an 18~nm$\cdot$rad double bend achromat (DBA) lattice: 
SPEAR~3. The low emittance design yields a high brightness beam, but
the stronger focusing in the DBA lattice increases chromaticity and 
beam sensitivity to machine errors. To ensure efficient injection and long
Touschek lifetime, an optimization of the design lattice and dynamic aperture 
has been performed. In this paper, we review the methods used to maximize 
the SPEAR~3 dynamic aperture including necessary optics modifications, 
choice of tune and phase advance, optimization of sextupole and
coupling correction, and modeling
effects of machine errors, wigglers and lattice 
periodicity.
\end{abstract}
\vfill
\begin{center}
{\it Presented at the 16th ICFA Beam Dynamics Workshop on Nonlinear and 
Collective Phenomena in Beam Physics, Arcidosso, Italy, 
September 1--5, 1998}
\end{center}
\newpage
\section*{Introduction}
\begin{figure}[b]
\includegraphics{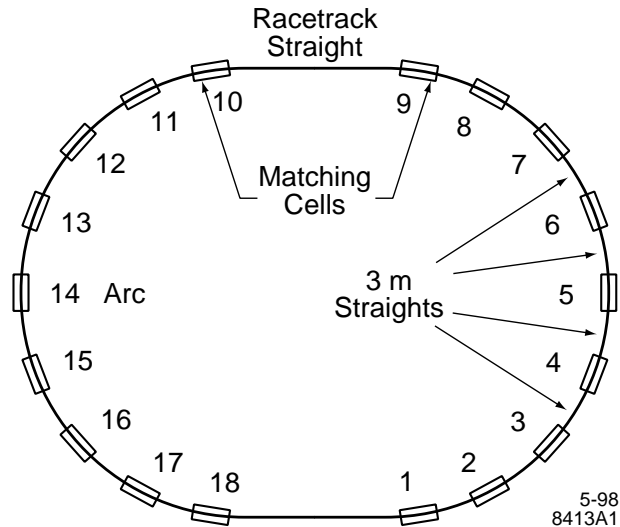}
\vspace{70mm}
\caption{Schematic of SPEAR tunnel girders.}
\label{y:f:girders}
\end{figure}

SPEAR~3 is the 3~GeV upgrade project under study at SSRL~\cite{y:upgrade}. 
It aims at replacing the current 130~nm$\cdot$rad FODO lattice with an
18~nm$\cdot$rad low emittance double bend achromat (DBA) lattice. To
minimize the cost of the project and to use the existing synchrotron light
beam lines, the new design~\cite{y:lattice,y:cdr} closely follows the racetrack 
configuration of the SPEAR tunnel, with the magnet positions fit to the 
18~magnet girders shown in Fig.~\ref{y:f:girders}. 
As in the current design, the 
SPEAR~3 lattice has two-fold symmetry and periodicity with two identical 
arcs and two long straight sections. Each arc in the new lattice has 
7~identical symmetric cells, and each straight section consists of two 
mirror symmetric matching cells.

\begin{figure}[t]
\includegraphics{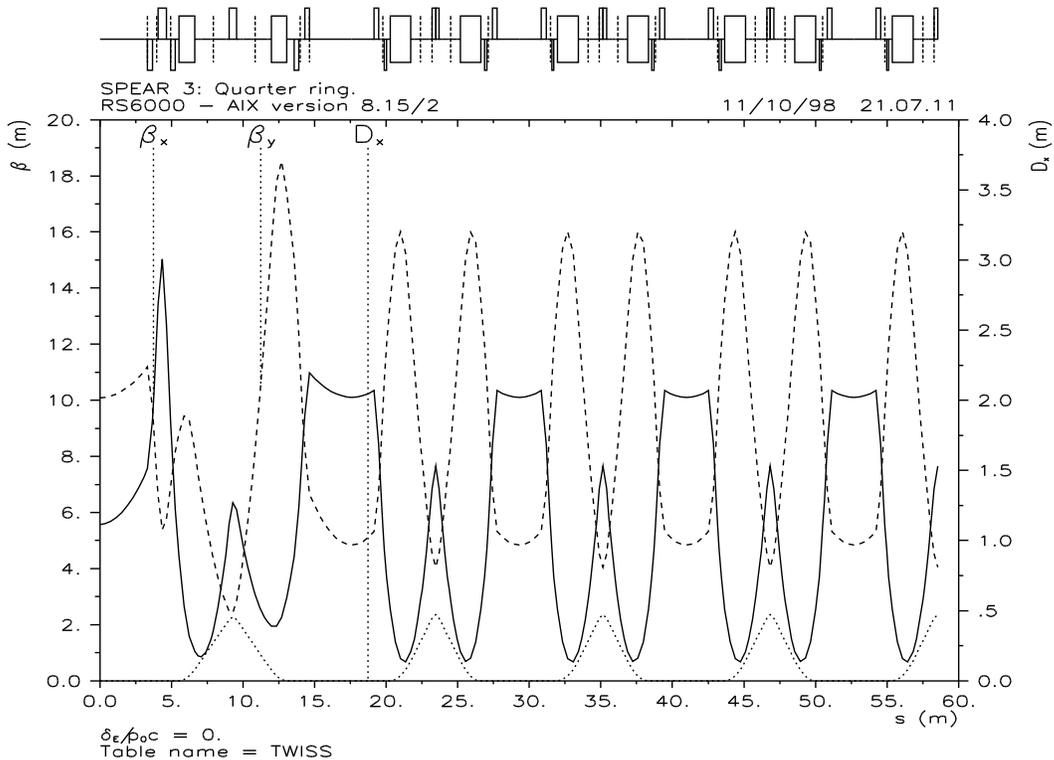}
\vspace{107mm}
\caption{Optics of One Quadrant of SPEAR~3 (from the center of the long
straight section to the middle of the arc).}
\label{y:f:qring}
\end{figure}

The lattice functions of one quarter of the ring are shown in 
Fig.~\ref{y:f:qring}.
The two bends and a quadrupole in the middle of each DBA cell compensate
the dispersion, while the two quadrupole doublets at each end control the
tune and cell $\beta$ functions. Since the new 
lattice cells have to fit to the existing 11.7~m cell length, it results in 
a compact DBA design, and hence increases the focusing. Similar to other 
light source lattices, it has been found 
advantageous to add vertical focusing to the 
bends~\cite{y:srs,y:srrc,y:trieste,y:cls} to relax 
the optics and reduce the strength of cell sextupoles by increasing the 
difference in their $\beta$~functions~\cite{y:grad}.
Each matching cell has an extra quadrupole
for a better optics matching, two 3/4 bends, with magnet positions adjusted
to maintain the ring circumference~\cite{y:jeff}
(the current 358.53~MHz RF~system 
will be used), and quadrupole strengths adjusted to
optimize $\beta$~functions and phase advance.

Though the DBA design has an advantage of a high brightness beam, 
its effect on the beam dynamics has to be verified. First, the 
lower emittance results in a higher particle density which increases 
probability of particle collisions inside the bunches and makes the Touschek 
effect the limiting factor for the beam lifetime. Secondly, the stronger 
focusing in the DBA lattice increases beam sensitivity to magnetic and 
chromatic errors and generates larger chromaticity. The latter requires 
strong sextupole correctors which increase the amplitude dependent 
and non-linear chromatic aberrations. These effects tend to reduce the 
dynamic aperture, and if the aperture is not sufficient for momentum errors
up to $\delta=3\%$ it will further reduce the Touschek beam lifetime. 
Adequate dynamic aperture is also important in order to minimize losses 
during horizontal oscillations of the injected beam.

Consequently, the low emittance design has to be optimized to achieve the
maximum dynamic aperture. It is especially important to maximize the 
horizontal size of dynamic aperture to minimize
the Touschek effect and allow large injection 
oscillations in SPEAR~3.
The improvement of dynamic aperture starts from optimization of linear
optics and correction systems. In the following sections we review the
modifications made to the initial
lattice design and present tracking studies 
including effects of magnetic and chromatic errors, perturbation due to
wigglers, and the effect of lattice periodicity.

All of the tracking study has been done using the 
LEGO code~\cite{y:lego} which
employs element-by-element tracking based on symplectic integration 
techniques~\cite{y:forest}. In a few cases we tested LEGO results against
other available tracking codes and found good agreement. In our study
we calculated dynamic aperture at injection point located at the
symmetry point between arc cells where $\beta_{x}=10.1$~m and
$\beta_{y}=4.8$~m. The other typical parameters were: number of tracking 
turns $N=1024$, linear chromaticity corrected to zero, and synchrotron 
oscillations included.
\vspace{5mm}
\section*{Error Free Dynamic Aperture}
Typically, the error free dynamic aperture serves as an upper limit for
the aperture with machine errors or with insertion devices (ID). It is 
therefore important to maximize first the dynamic aperture for the ideal 
lattice without any magnetic or misalignment errors. Maximizing the error
free dynamic aperture necessarily involves optimization of linear 
optics and the chromaticity correction system, minimization of chromatic and 
high order effects, and optimizing the betatron tune.
\subsection*{Cell Optics}
The DBA cell optics was made to fit the existing 11.7~m cell length with 
the magnet positions constrained to provide $\sim 3$~m space for the insertion 
devices and with bend positions kept to fit to the current synchrotron light 
beam lines. This results in a compact DBA design which leads to stronger 
focusing and, hence, increased beam sensitivity to machine errors. Though 
the achromat lattice eliminates dispersion in the insertion devices and
at injection, it limits the available positions for chromatic sextupoles 
to rather short dispersive regions between the bends and the middle 
quadrupole QFC. The close proximity of the SF and SD sextupoles reduces 
their effectiveness and requires larger strengths.

To relax the cell optics it has been found advantageous to add vertical 
focusing to the cell bends. This results in a better separation of 
horizontal and vertical focusing and reduces the quadrupole strength in 
the doublets. Most importantly, due to increased $\beta_{y}$ at SD
sextupoles it provides better optical separation between the SF and SD
and reduces their strength. To further reduce the strengths of the
sextupoles and QFC quadrupole, the bends were placed as far from the 
cell center as it is possible within existing geometric constraints.
\subsection*{Working Tune and Phase Advance}
The choice for the phase advance in the arc cells was to be near 
$\mu_{x}\approx 0.75\times 2\pi$ and $\mu_{y}\approx 0.25\times 2\pi$. 
This provides favorable conditions for local cancellation of: 
1)~geometric aberrations from arc sextupoles located $-I$ apart, and 
2)~first order chromatic beta waves from sextupoles and quadrupoles 
located $90^{o}$ apart, as well as systematic quadrupole errors. It is 
worth to note that the high horizontal phase advance is due to the achromat 
design which requires $\mu_{x}=\pi$ between the bends.

Initially, the matching cells were designed to provide {\it{I}}-transformation
between the two arcs in both planes~\cite{y:garren}. This would give the
advantage of having effectively a 14-period lattice since the matching cells 
would be virtually invisible in the first order to the on-momentum 
particles. Generally, the high periodicity optics provides better 
cancellation of systematic errors. 

With the above choices the total tune would be near $\nu_{x}\approx 14.5$ 
and $\nu_{y}\approx 5.5$. To move the working tune away from the half 
integer resonance, the phase advance in arc and/or matching cells has to 
be adjusted. To minimize the resistive wall impedance effects~\cite{y:reswall}
it is favorable to move the tune into the lower quarter on the tune plane 
($\nu < 1/2$). Tracking studies showed that relaxing the 
phase advance through the matching cells 
improves the SPEAR~3 off-momentum dynamic aperture since it reduces the 
chromaticity and the strengths of matching quads. On the other hand, 
relaxing the arc cells would increase the arc sextupole strengths because 
of unfavorable change of the cell $\beta$~functions at the sextupoles.

The optimum phase conditions have been identified by performing a horizontal 
dynamic aperture scan across the matching cell phase advance $\mu_{x}$ 
and $\mu_{y}$. The on-momentum and off-momentum dynamic aperture were maximized 
at about $\mu_{x}=0.78\times 2\pi$ and $\mu_{y}=0.42\times 2\pi$ per 
matching cell.

\begin{figure}[tb]
\includegraphics{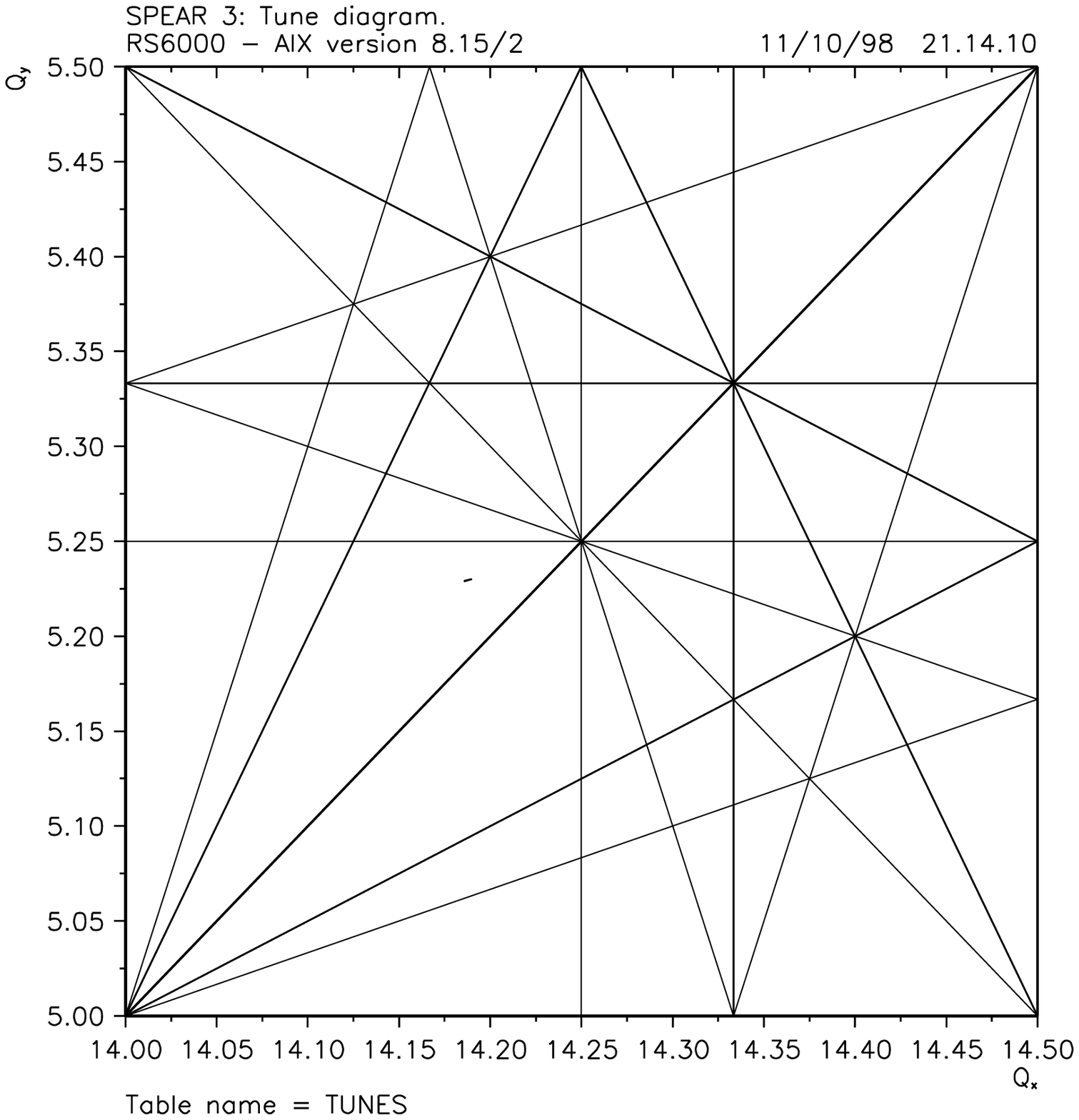}
\vspace{119mm}
\caption{SPEAR~3 Working Tune on the Tune Plane.}
\label{y:f:tune}
\end{figure}

To minimize the effect of strong low order betatron resonances the 
location of the working tune on the tune plane has been chosen slightly
below~.25, away from the 3rd and 4th order resonance lines. The final 
choice ($\nu_{x}=14.19$, $\nu_{y}=5.23$) was based on favorable 
horizontal injection conditions and the results of dynamic aperture tune 
scan. The two dimensional diagram of horizontal dynamic aperture (in 
number of $\sigma_x$) versus $x$ and $y$ tune is shown in Table~1,
where $\sigma_{x}\approx 0.45$~mm is the horizontal rms beam size at
injection point. During the 
aperture scan the tune was varied by changing arc phase advance, and the 
lattice was kept matched at all tunes. The above scan also 
included a set of random machine errors which will be described in the 
following sections. With the chosen tune, the phase advance per arc cell is 
$\mu_{x}=0.7907\times 2\pi$ and $\mu_{y}=0.2536\times 2\pi$. The location 
of the working tune on the tune plane along with betatron resonance lines 
up to 4th order is shown in Fig.~\ref{y:f:tune}.

\begin{table}[tb]
\caption{Horizontal Dynamic Aperture (in number of $\sigma_{x}=0.45$~mm) 
versus $\nu_{x}$, $\nu_{y}$. The Tracking Included a Set of Random Machine
Errors and 1\% Momentum Error.}
\label{y:t:tunscan}
\begin{center}
\begin{tabular}{lccccccccccccccc}
\boldmath{$\nu_{x}\rightarrow$}
 &.16&.17&.18&.19&.20&.21&.22&.23&.24&.25&.26&.27&.28&.29&.30\\
\boldmath{$\downarrow\nu_{y}$}
 & & & & & & & & & & & & & & & \\
\tableline
.30&49&49&49&49&51&46&47&47&46&44&41&38&36&49&48\\
.29&51&46&52&51&48&46&47&47&47&44&41&39&46&49&48\\
.28&49&49&50&51&49&48&51&49&46&44&42&40&36&48&48\\
.27&50&49&49&51&50&48&46&49&46&42&40&40&37&38&37\\
.26&49&46&47&51&49&46&44&47&43&42&42&41&41&39&41\\
.25&47&47&49&51&49&48&44&43&46&45&42&42&37&41&49\\
.24&47&47&46&50&47&49&45&44&48&46&43&41&45&43&50\\
.23&48&49&46&49&48&48&46&50&48&45&44&49&47&42&42\\
.22&48&48&45&44&49&48&50&48&48&45&46&42&42&43&43\\
.21&49&45&46&46&48&50&50&51&48&46&44&44&46&46&44\\
.20&48&48&44&46&41&47&44&50&48&45&46&42&47&48&45\\
.19&48&46&45&50&48&45&47&49&46&47&47&48&47&47&45\\
.18&47&45&47&51&49&47&47&42&46&47&43&49&46&43&40\\
.17&53&49&48&49&50&48&47&45&45&43&46&40&45&44&42\\
.16&51&49&48&50&49&50&47&47&44&42&41&36&42&41&40
\end{tabular}
\end{center}
\end{table}

\subsection*{Chromatic Correction}
Efficient chromatic correction is essential for a large off-momentum
dynamic aperture and for a long Touschek lifetime. Since the sextupoles
also give rise to non-linear geometric aberrations it is important to
minimize these effects by using compensation techniques and reducing the
sextupole strengths.

As mentioned previously, the choice for phase advance in the arc cells 
provides conditions for local compensation of sextupole geometric aberrations
and chromatic beta waves from arc quadrupoles and sextupoles. This scheme 
would work optimally if the number of arc cells was $4\times \it{integer}$.
With 7~arc cells in the design, however, this correction is not complete, 
and the geometric constraints do not allow for 8 identical arc cells.

The chromaticity correction using only two sextupole families in the arc 
cells does not provide adequate dynamic aperture for off-momentum 
particles. Since the two family sextupoles can only compensate the linear 
chromaticity, the off-momentum aperture is mostly affected by the non-linear 
chromatic effects. The reason for the large high order chromaticity is due 
to the matching cells which contribute about 20\% to the total chromaticity 
and break periodicity of the 14 arc cells. Two additional families of 
sextupoles (SFI,~SDI) placed in the matching cells, similar to the arc cells, 
help to reduce the non-linear terms and significantly improve the 
off-momentum dynamic aperture. Table~2 compares 
HARMON~\cite{y:mad} calculations of the high order chromaticity for the 
lattice with and without SFI,~SDI sextupoles.

\begin{table}[htb]
\caption{2nd and 3rd Order Chromaticity for SPEAR~3 with and w/o
Matching Cell Sextupoles.}
\label{y:t:chrom}
\begin{center}
\begin{tabular}{lcccc}
\textbf{2nd and 3rd Order} & \boldmath{$\underline{d\nu_{x}}$} 
                           & \boldmath{$\underline{d\nu_{y}}$} 
                           & \boldmath{$\underline{d\nu_{x}}$} 
                           & \boldmath{$\underline{d\nu_{y}}$} \\
\textbf{Chromaticity} & \boldmath{$d\delta^2$} 
                      & \boldmath{$d\delta^2$} 
                      & \boldmath{$d\delta^3$} 
                      & \boldmath{$d\delta^3$} \\
\tableline
w/o SFI/SDI & -117 & -52 & -674 & -301 \\
with SFI/SDI & -48 & -12 & -228 & 74
\end{tabular}
\end{center}
\end{table}

The matching cell sextupoles also generate geometric aberrations and 
therefore have to be kept relatively weak in order to preserve the 
on-momentum aperture. The optimum strengths of the matching cell 
sextupoles have been evaluated by performing a horizontal aperture scan 
versus SFI,~SDI strengths while using the arc sextupoles to keep the ring 
chromaticity constant. Other chromatic correction schemes which use 
more sextupole families in the arcs did not result in a better aperture.

As it was mentioned earlier, the close proximity of the SF,~SD sextupoles in
the short dispersive region between bends and QFC quadrupole reduces the 
effectiveness of the sextupoles and increases their strength. To
increase the optical separation of the sextupoles two other
options were studied. In one option, the SD sextupole was moved away from
the SF by combining with part of the adjacent bend. This increased the 
$\beta_y$~function at the SD, but the dispersion inside the bend was rather 
low. The tracking study showed that the dynamic aperture reduces in this 
option. In the second study, the SF sextupole was combined with the QFC 
quadrupole. Due to higher dispersion and $\beta_x$ at the QFC, this led 
to weaker SF sextupoles and potentially reduced high order sextupole effects. 
The dynamic aperture, however, did not improve using this option.
Consequently, in the current design the sextupoles are kept separate from 
the bends and QFC quadrupole.

\begin{figure}[tb]
\includegraphics{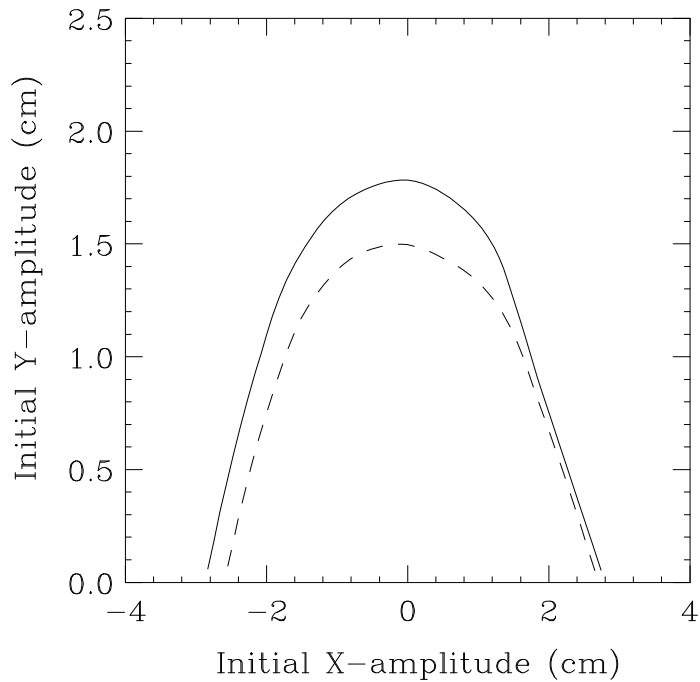}
\vspace{95mm}
\caption{Error Free Dynamic Aperture for on-Momentum (solid) and
$\delta=3\%$ off-Momentum (dash) Particles.}
\label{y:f:bare}
\end{figure}

The resultant error free dynamic aperture for on-momentum and $\delta=3\%$ 
particles is shown in Fig.~\ref{y:f:bare}. The axes refer to the initial 
particle amplitude at the 
injection point, and the two curves show the boundary 
for the stable motion for on and off-momentum particles. Clearly, the 
off-momentum aperture is very robust against momentum errors and 
provides favorable conditions for a long Touschek lifetime.

\section*{Effect of Machine Errors}
Magnetic field and alignment errors introduce optics perturbations and
enhance effects of resonances that limit dynamic aperture. For conservative
results, we included several different classes of magnet errors in the 
tracking studies including random main field errors, random and systematic 
multipole errors, and random alignment errors. Since the skew quadrupoles 
combined with sextupoles in SPEAR~3 generate a skew octupole field, 
and the large orbit variation in the $10.6^{\circ}$ bend samples 
high order multipole fields, 
these effects were added to the error set. The effect of positive linear
chromaticity, large $\beta$ function distortions, orbit distortions, 
large amplitude coupling and insertion devices were studied. Harmonic 
sextupole correction has been tried to improve the dynamic aperture, 
and the effect of lattice periodicity was analyzed.
\subsection*{Alignment and Field Errors} 
The magnet error and alignment 
specifications used for the SPEAR~3 tracking studies can be
met with standard manufacturing techniques. For 
tracking simulations, the following values for rms errors were used with 
$2\sigma$ truncation to simulate realistic quality control.
\subsubsection*{Alignment}
The alignment rms errors shown in Table~3 can be achieved with survey
techniques used in practice and are large enough to yield conservative 
tracking results. The  dipole alignment specification is the same as 
for quadrupoles since the SPEAR~3 dipoles include a strong quadrupole field.

\begin{table}[htb]
\caption{Alignment rms Errors Used in Tracking Studies.}
\label{y:t:align}
\begin{center}
\begin{tabular}{lccc}
\textbf{Element} & \boldmath{$\Delta x$~($\mu m$)} 
                 & \boldmath{$\Delta y$~($\mu m$)} 
                 & \textbf{Roll}~\boldmath{($\mu rad$)} \\
\tableline
Dipole & 200 & 200 & 500 \\
Quadrupole & 200 & 200 & 500 \\
Sextupole & 200 & 200 & 500
\end{tabular}
\end{center}
\end{table}
\subsubsection*{Systematic Multipole Errors}
Systematic multipole field errors are field components of higher order than 
the main field which apply to all magnets with a common core design.
In LEGO, the multipole errors are defined in terms of a ratio of the 
multipole field $\Delta B_{n}$ (normal or skew) to the main magnet field 
$B$ at radius $r$, where $n=1,2,3,\ldots$ is the multipole order starting 
with a bend. The normal rms values $\Delta B_{n}/B$ used for the SPEAR~3 
magnets are listed in Table~4. For the quadrupole magnets, only the allowed 
multipoles were used. For the gradient dipole magnets and sextupoles combined 
with skew quads, all systematic multipoles were used with the largest values
shown in the Table~4.

\begin{table}[htb]
\caption{Systematic rms Multipole Field Errors.}
\label{y:t:sysmult}
\begin{center}
\begin{tabular}{lccc}
\textbf{Magnet} & \boldmath{$r(mm)$} 
                & \boldmath{$n$} 
                & \boldmath{$\Delta B_{n}/B$} \\
\tableline
Dipole & 30 & 2 & $1\times 10^{-4}$ \\
       & & 3-14 & $5\times 10^{-4}$ \\
Quadrupole & 32 & 6,10,14 & $5\times 10^{-4}$ \\
Sextupole & 32 & 4 & $-8.8\times 10^{-4}$ \\
          & & 5 & $-6.6\times 10^{-4}$ \\
          & & 9 & $-1.6\times 10^{-3}$ \\
          & & 15 & $-4.5\times 10^{-4}$
\end{tabular}
\end{center}
\end{table}

In the SPEAR~3 design the skew quadrupoles used for coupling correction 
are combined with sextupoles. Such combined magnets also generate a skew 
octupole field proportional to the skew quad strength. At radius of 
32~mm the magnitude of the octupole field will be 57\% of the skew quad field. 
In the tracking, the skew octupole component was systematically added in 
proportion to the skew quadrupole field. It is worth to note that typical
skew quad strengths required for coupling correction are less than
1\% of the main ring quadrupole strengths.
\subsubsection*{Random Field Errors}
Differences in magnetic core length will give rise to $\sim 10^{-3}$ random 
main field errors. Normal random multipole errors, introduced by magnet 
assembly imperfections, are listed in Table~5. To achieve conservative 
tracking results, large values were specified for the 
random $n=$3,6,10,14 multipoles on the quadrupole magnets.
\begin{table}[htb]
\caption{Random rms Multipole Field Errors.}
\label{y:t:ranmult}
\begin{center}
\begin{tabular}{lccc}
\textbf{Magnet} & \boldmath{$r(mm)$}
                & \boldmath{$n$} 
                & \boldmath{$\Delta B_{n}/B$} \\
\tableline
Dipole & 30 & 2 & $1\times 10^{-4}$ \\
Quadrupole & 32 & 3,6,10,14 & $5\times 10^{-4}$ \\
           &    & 4,5,7-9,11-13 & $1\times 10^{-4}$ \\
Sextupole & 32 & 5 & $1.5\times 10^{-3}$ \\
          & & 7 & $4.8\times 10^{-4}$
\end{tabular}
\end{center}
\end{table}
\subsection*{Coupling Correction}
The initial design of coupling correction employed four independent skew
quadrupoles placed in non-dispersive regions in the matching cells.
The four skew quads were used to uncouple the $4\times 4$ one turn
transfer matrix. However, the skew quad placement in the matching cells
did not
provide enough variation of the sum and difference phase advance
($\mu_{x}\pm \mu_{y}$) for efficient orthogonal
correction. Depending on the set of random machine errors, this occasionally
led to strong skew quadrupoles and reduced dynamic aperture.

Further study showed that it was beneficial for dynamic aperture to use skew
quadrupole components on the chromatic sextupoles located
in the dispersive regions of the arc cells~\cite{y:cdr}.
This configuration provided
a more orthogonal set of the skew quad positions with reduced
strengths. The negative effects, such as induced vertical dispersion,
were small compared to improved aperture and robustness of the correction.
In total, 24 skew quads were arranged in
four families and placed at their optimum phase positions. One other negative
effect is that a skew quadrupole in a sextupole magnet gives rise to a 
systematic skew octupole field. In our tracking study, the effect of this 
field did not reduce the dynamic aperture. As a future option, the large 
number of skew quads in the above scheme allows
expansion of the number of 
independent families with correction of the vertical dispersion as 
well.
\subsection*{Dynamic Aperture with Errors}
For a realistic simulation with errors, LEGO first generates and
adds the chosen set of errors to the magnets, then iteratively applies 
correction schemes to minimize the optics perturbation, and finally tracks 
particles with a variety of initial amplitudes to define the dynamic aperture.
The basic set of correction schemes in LEGO includes tune, orbit, linear
chromaticity and coupling correction systems. In the SPEAR~3 simulation, the 
tune was corrected by using two families of doublet quads in the arc cells. 
The nominal orbit correction routine in LEGO is based on a three corrector 
bump scheme, but other techniques can be implemented as well. Typically in
this study, the linear chromaticity was adjusted to zero with the two 
families of sextupoles in the arc cells, while the matching cell
sextupole strength was kept constant. The coupling correction was done 
by using the four family skew quad correction scheme described in the 
previous section. An RF~voltage of 3.2~MV was used to generate synchrotron 
oscillations for off-momentum particles.

\begin{figure}[tb]
\includegraphics{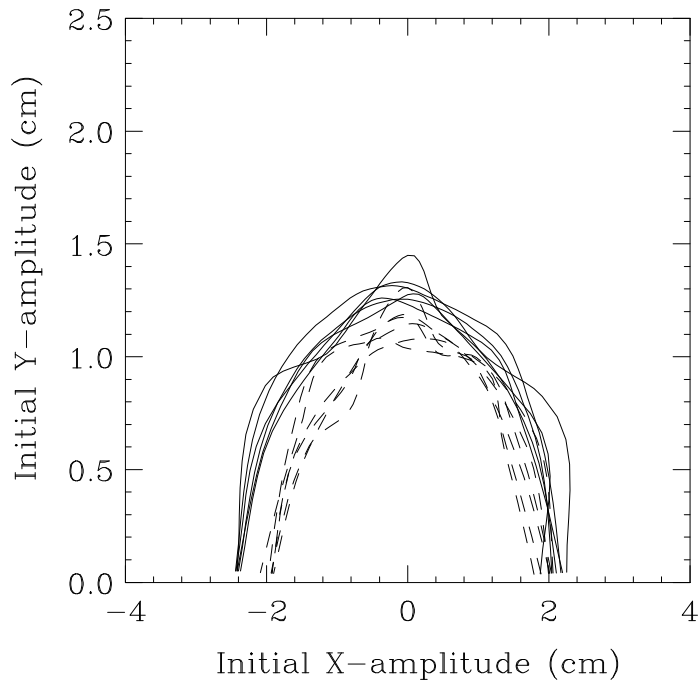}
\vspace{95mm}
\caption{Dynamic Aperture for 6 Seeds of SPEAR~3 Machine Errors
for $\delta=0$~(solid) and 3\%~(dash) Momentum Oscillations.}
\label{y:f:seed}
\end{figure}

The resultant dynamic aperture for 6 random seeds of machine errors for 
on-momentum and $\delta =3\%$ off-momentum particles with the described 
correction schemes is shown in Fig.~\ref{y:f:seed}. The linear chromaticity 
was set to zero in this case. Compared to Fig.~\ref{y:f:bare} (no errors),
the dynamic aperture has reduced by about 20-30\%. Since insertion 
devices were not included in this 
calculation, this reduction is solely due to the machine errors and quality 
of correction procedures described above. As in the case of the error free 
lattice, the off-momentum aperture for machine with errors is comfortably
large. As the Fig.~\ref{y:f:seed} shows, the horizontal dynamic aperture 
is in the range 
of 18 to 20~mm for all particles within $\delta=\pm 3\%$ momentum range. 
This provides favorable conditions for a long Touschek lifetime and
sufficient room for horizontal injection oscillations.
\subsection*{Positive Linear Chromaticity}
Though most of this study was done with the linear chromaticity corrected to
zero, in real machines the value of $\xi=\Delta\nu/\delta$ is typically set 
slightly positive, up to several units, to avoid effects such as head-tail
instability. The main impact of the positive chromaticity on the dynamic
aperture is from the increased tune spread in the beam. Due to synchrotron
oscillations the particles with large momentum errors would sample a larger
area on the tune plane and might cross more harmful betatron resonances.
As a result, the momentum aperture can be significantly reduced if the
linear chromaticity is too large. The effect on the on-momentum aperture is 
typically smaller due only to the increased strength of sextupole correctors.

\begin{figure}[ptb]
\includegraphics{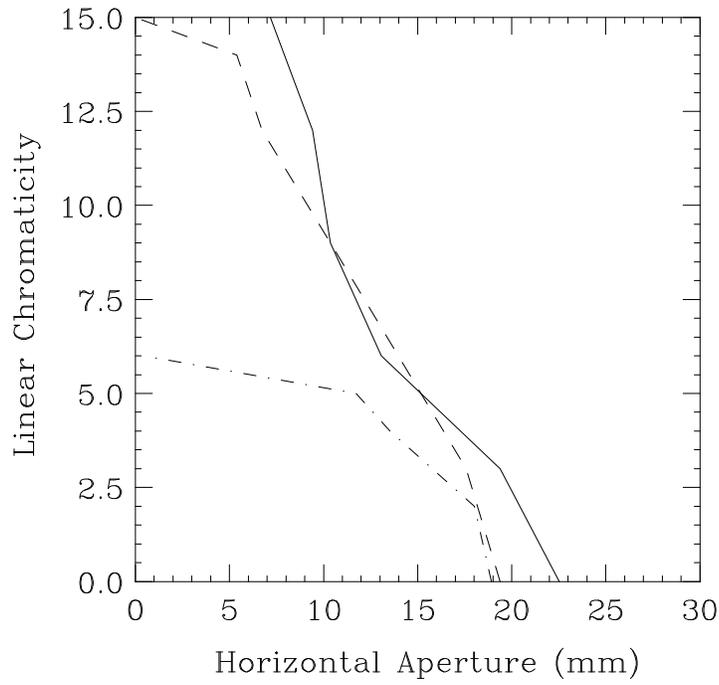}
\vspace{95mm}
\caption{Horizontal Dynamic Aperture versus Linear Chromaticity for
$\delta=0$~(solid), 1\%~(dash) and 3\%~(dot-dash) Momentum Oscillations.}
\label{y:f:axchrom}
\end{figure}

\begin{figure}[ptb]
\includegraphics{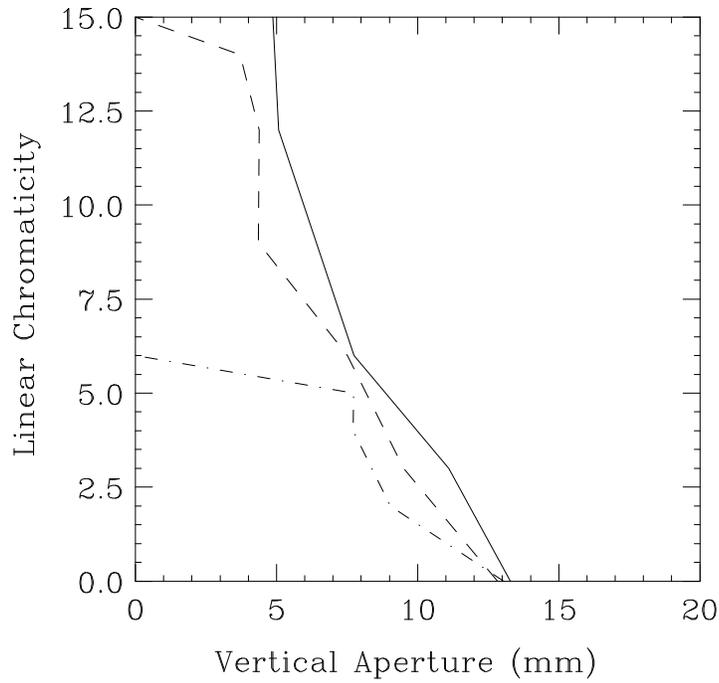}
\vspace{95mm}
\caption{Vertical Dynamic Aperture versus Linear Chromaticity for
$\delta=0$~(solid), 1\%~(dash) and 3\%~(dot-dash) Momentum Oscillations.}
\label{y:f:aychrom}
\end{figure}

Fig.~\ref{y:f:axchrom},\ref{y:f:aychrom} show dependence of linear chromaticity 
on the dynamic aperture for the particles with momentum oscillations of 
$\delta=0,1\%$ and 3\%. The tracking included a full random set of machine 
errors, and the chromaticity was set equal in the $x$ and $y$ planes.
Fig.~\ref{y:f:axchrom},\ref{y:f:aychrom} show that the particles with 
momentum errors of $\delta=1\%$ and 3\% lose stability 
at $\xi\approx 15$ and 6, respectively.
Clearly, this is the effect of a half integer resonance. For the SPEAR~3
working tune ($\nu_{x}=14.19$, $\nu_{y}=5.23$) the off-momentum particles
would likely be lost when momentum dependent tune shift approaches to
$\Delta\nu\approx -0.2$. 
The dynamic aperture for the core beam (low momentum error) is not 
significantly reduced even for the large positive chromaticity.
However, the beam lifetime can be reduced for
$\xi>5$ due to Touschek effect, since the scattered particles with large 
$\delta$ may not survive.

\subsection*{Feed-down Studies}
Since the electron beam in a SPEAR~3 dipole magnet follows a $10.6^{\circ}$
arc orbit with $\pm16.6$~mm sagitta, even the particles traveling along 
the ideal trajectory will sample the full set of high order multipole field. 
Taking into account the large horizontal excursion of the beam orbit, the 
dipoles are specified to have a full 92~mm good field region~\cite{y:cdr}. 
Since normally the Taylor expansion of the multipole field is specified 
around the central magnetic axis, each individual multipole of this field 
expanded about the curved beam orbit in a dipole will generate (feed-down) 
an additional full set of lower order multipoles proportional to the orbit 
displacement. Both the nominal and feed-down multipoles have to be combined 
to realistically estimate the effect of dipole errors. Though this feed-down 
effect might be negligible for large machines, it has to be verified for 
smaller machines with large sagitta in the dipoles.

In LEGO and most other codes, the multipole field in a dipole would be 
expanded with respect to the ideal orbit, not the magnetic axis. Therefore,
when applying the multipole errors in a dipole, the feed-down terms were
included in addition to the nominal set of multipoles specified around 
the magnetic center. Since the feed-down terms depend on the orbit 
displacement, we used the following technique to evaluate this effect.
The 1.45~m dipole was `sliced' in a reasonably short pieces and the average 
orbit displacement was calculated for each slice. Based on the orbit
displacement in each slice and the nominal set of multipole fields, the 
systematic feed-down terms were calculated for each slice. The tracking
simulation was then done including sliced dipoles with multipole feed-down
terms. The results showed no degradation of the dynamic aperture 
due to this effect.

Since the sliced dipoles would significantly increase the computer
time for element-by-element tracking, normally this model was not used
in tracking studies. However, the magnitude of the nominal dipole multipole 
field used in tracking runs without explicit feed-down effects
was set conservatively large to produce comparable field around
the orbit.
\subsection*{Large Beta Distortion}
In the tracking, typical $\beta$ distortions caused by the SPEAR~3 
specification errors after correction were on the order of 
$\Delta\beta/\beta=\pm(5-10)\%$. However, in a real machine it is not unusual 
to observe much larger modulations since some of the design specifications 
may not be achieved, especially during commissioning. The $\beta$ distortions
lead to a larger beam size and may increase the effects of high order field 
and resonances.

\begin{figure}[tb]
\includegraphics{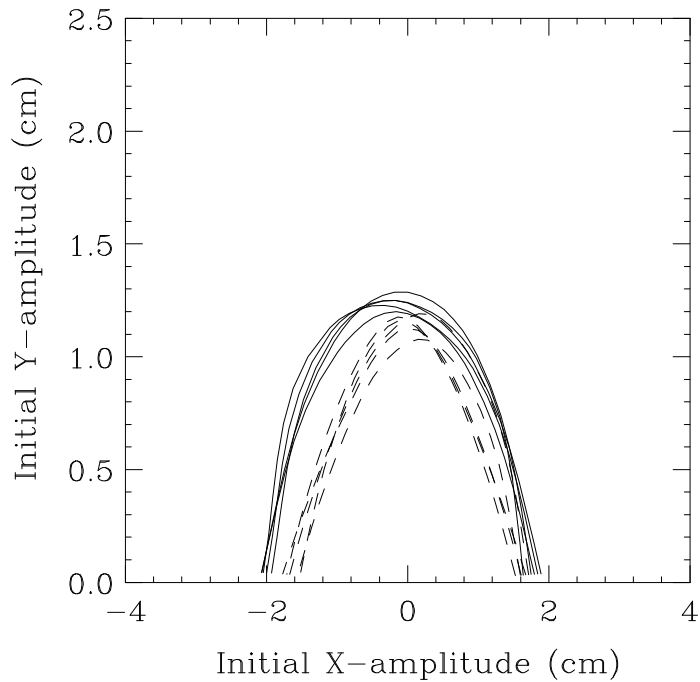}
\vspace{95mm}
\caption{Dynamic Aperture with Large $\beta$ Distortion
($\pm30\%$~(x), $\pm20\%$~(y)) for on-Momentum (solid) and $\delta=3\%$
off-Momentum (dash) Particles for 5 Random Error Seeds.}
\label{y:f:bbeat}
\end{figure}

To verify the effect of large $\beta$ modulation on the SPEAR~3 dynamic 
aperture, the quadrupole errors in two quad families in the matching 
cells were increased to the level of a few percent to produce 
$\Delta\beta_{x}/\beta_{x}\approx\pm30\%$ and
$\Delta\beta_{y}/\beta_{y}\approx\pm20\%$. The calculated dynamic aperture
for 5 seeds of random machine errors is shown in Fig.~\ref{y:f:bbeat}. 
The average reduction of the aperture due to this $\beta$ distortion is 
about 15\% compared to Fig.~\ref{y:f:seed}. Though this aperture is 
still adequate to operate the
machine, it is clear that such large quadrupole errors have to be
identified during initial operation and corrected.

\subsection*{Orbit Distortion}
Large orbit distortions can cause a reduction of dynamic aperture since the 
electron beam would experience much larger effects from non-linear sextupole
and multipole fields. Normally, in this study, the rms orbit was corrected
very well, down to the level of $\sim 100\mu m$. To investigate large orbit 
effects, an additional set of magnet alignment errors was introduced in
LEGO just prior to tracking. Using additional uncorrected random 
alignment errors with $\sigma_{\Delta x}=80\mu m$, 
$\sigma_{\Delta y}=40\mu m$, and $50\mu rad$ roll, rms orbit distortions 
of up to 3~mm in the horizontal and 1.5~mm in the vertical plane were 
generated. In each case, the on-energy dynamic aperture was reduced by 
a maximum of 2~mm in the horizontal and vertical planes for 6 
different seeds. The off-energy aperture was reduced similarly 
compared to the off-energy dynamic aperture without orbit distortions.

For safe machine operation, the peak vertical orbit excursion must be 
held to $<1$~mm and the horizontal excursion should be $<3$~mm, so orbit 
induced reduction of dynamic aperture is negligible. Absolute BPM reading
errors are expected to be on the order of a few hundred $\mu m$
or less. The low sensitivity of the aperture to orbit 
distortions should simplify initial injection and machine commissioning 
at low currents.

In a related test, we studied large sextupole misalignment while the orbit
was well corrected. The sextupole misalignments generate random
quadrupole errors and result in optics distortion and coupling. To simulate
this effect, rms sextupole misalignments of 1~mm were assigned in both 
planes, and the orbit was corrected down to a few hundred $\mu m$ level. 
The 1~mm sextupole displacements generate an order of magnitude larger 
quadrupole errors compared to the specified field errors in the ring 
quadrupoles. Of the 6 seeds studied, 5 cases showed 
$>17$~mm horizontal dynamic aperture
for on-momentum particles. The vertical aperture was always larger than 
the $\pm 6$~mm ID chamber size. At $\delta=3\%$, the horizontal aperture 
remained above 13~mm. The one 'bad' seed produced 13~mm horizontal dynamic 
aperture on-momentum and about the same result for 3\% off-momentum particles.
In practice, with sextupole alignments much better than 1~mm rms, no reduction
of dynamic aperture is expected. 
\subsection*{Large Amplitude Coupling}
In order to monitor the full extent of the dynamic aperture, the SPEAR~3 
tracking simulations did not include physical apertures. In practice, 
however, the vacuum chamber has horizontal and physical apertures that 
can limit beam lifetime. In the vertical plane, for example, SPEAR has two
insertion devices with $y=\pm6$~mm vacuum chambers that define the 
vertical acceptance. Although the height of the ID chambers yield 
acceptable gas scattering lifetime~\cite{y:cdr}, they can limit the 
Touschek lifetime if strong coupling is present. In the presence of 
machine errors and strong sextupole fields, for instance, particles 
with large horizontal amplitudes can reach resonances which couple the
horizontal motion into the vertical plane. This effect has been observed 
in operational machines~\cite{y:robin}.

\begin{figure}[tb]
\includegraphics{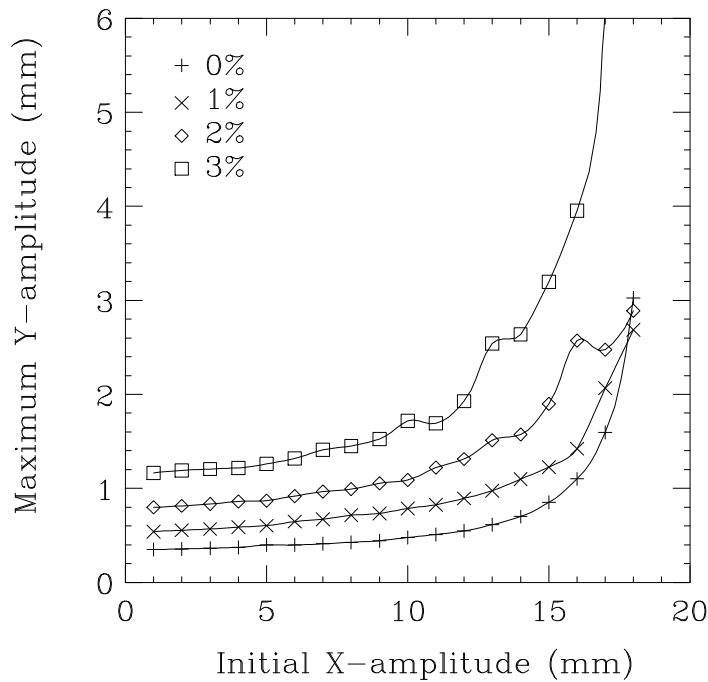}
\vspace{95mm}
\caption{Peak Vertical Excursion as a Function of Initial Horizontal
Amplitude for $\delta=0,1,2,$ and 3\% Momentum Oscillation. Results
Averaged over 6 Seeds of Machine Errors.}
\label{y:f:xycoupl}
\end{figure}

To study the effect of large amplitude coupling, we launched particles 
with variable horizontal and synchrotron oscillation amplitudes and 
monitored the maximum vertical excursion. Fig.~\ref{y:f:xycoupl} 
shows the degree of $x$-$y$ coupling  as a function 
of initial horizontal amplitude for particles 
with $\delta=0,1,2,$ and 3\% energy oscillations. Each particle was 
launched with a small initial vertical amplitude of $100\mu m$ and the peak
vertical amplitude was monitored for 1024~turns at the ID location. 
The plot shows the average value taken over 6 machines with different 
error seeds. Based on these results, one can conservatively anticipate 
an effective reduction of horizontal aperture from 20~mm to about 18~mm, 
and a corresponding reduction in Touschek lifetime from $\approx 135$~hrs
to $\approx 125$~hrs~\cite{y:cdr}. The 10~mm injection oscillations in
the horizontal plane should not be effected by this coupling.

\subsection*{Harmonic Sextupoles}
The first non-linear field magnets typically introduced in the optics 
are sextupoles which are placed in dispersive regions to correct
chromaticity. In addition to their chromatic effect, 
sextupoles generate geometric 
aberrations such as amplitude dependent tune shift and high order 
resonances. Even without machine errors these effects may significantly 
limit dynamic aperture. Therefore, it is important to verify this 
limitation and use available techniques to minimize a reduction of 
aperture.

Though the choice of cell phase advance in the SPEAR~3 arcs helps
to reduce the sextupole geometric aberrations, they are not completely 
canceled. To further minimize these effects, additional `harmonic' 
sextupoles can be used~\cite{y:hsex}. These sextupoles are usually placed 
in non-dispersive regions to avoid their chromatic effect, and their 
strengths are optimized to minimize the total amplitude dependent tune shift.
Since this tune shift can be mathematically decomposed into a series 
of tune harmonic components generated by sextupoles~\cite{y:hsex}, 
one technique to reduce the tune
shift is to minimize the strongest harmonic components.

\begin{figure}[tb]
\includegraphics{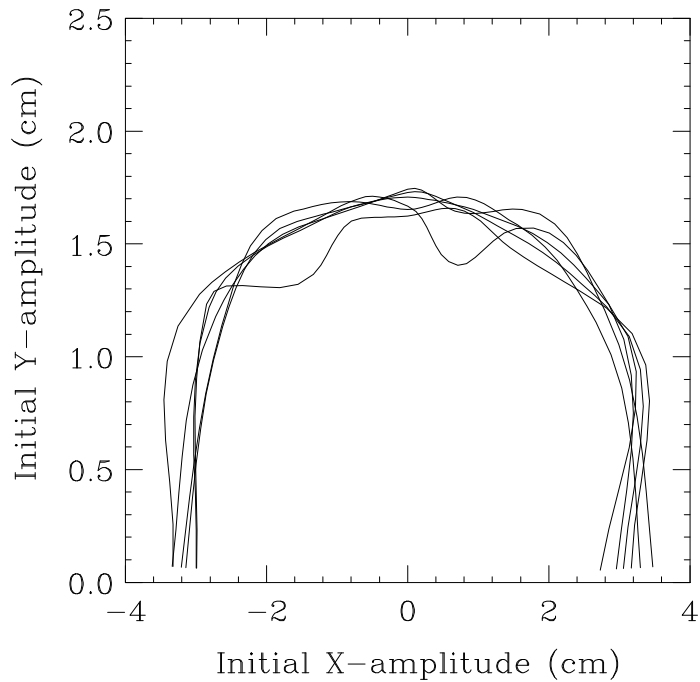}
\vspace{95mm}
\caption{On-Momentum Dynamic Aperture without Sextupoles
for 6 Random Error Seeds with RF Cavities Turned off.}
\label{y:f:nosex}
\end{figure}

Obviously, the upper limit for dynamic aperture with sextupoles is
the aperture where sextupole aberrations are not present. One way to
evaluate this limit is to track on-momentum particles in the lattice with
sextupoles turned off. The RF cavities have to be turned off as well to
eliminate any chromatic effects. The result of this tracking for SPEAR~3
with 6 random seeds of machine errors is shown in Fig.~\ref{y:f:nosex}. 
It follows that if the sextupoles are perfectly compensated the 
horizontal aperture could be as large as 28~mm or 40\% larger 
compared to Fig.~\ref{y:f:seed}. Realistically,
this limit may not be achieved since all sextupole aberrations have to 
be canceled all at once, and even with perfect global compensation the 
local effects would be present. Effectiveness of a harmonic sextupole 
system would also
depend on the phase advance of a particular optics configuration.

\begin{figure}[ptb]
\includegraphics{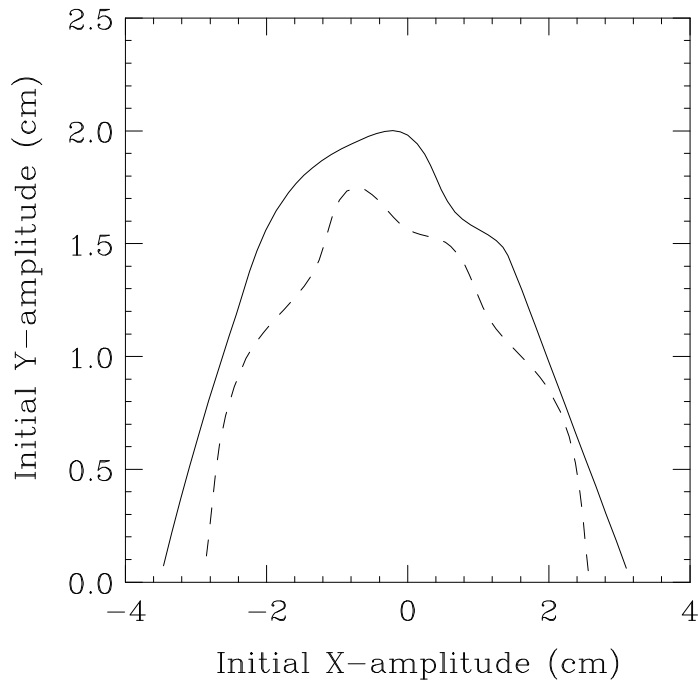}
\vspace{95mm}
\caption{Error Free Dynamic Aperture with Two Family Harmonic
Sextupoles in Arc Cells for 0~(solid) and 3\%~(dash) Momentum 
Oscillations.}
\label{y:f:hsxbare}
\end{figure}

\begin{figure}[ptb]
\includegraphics{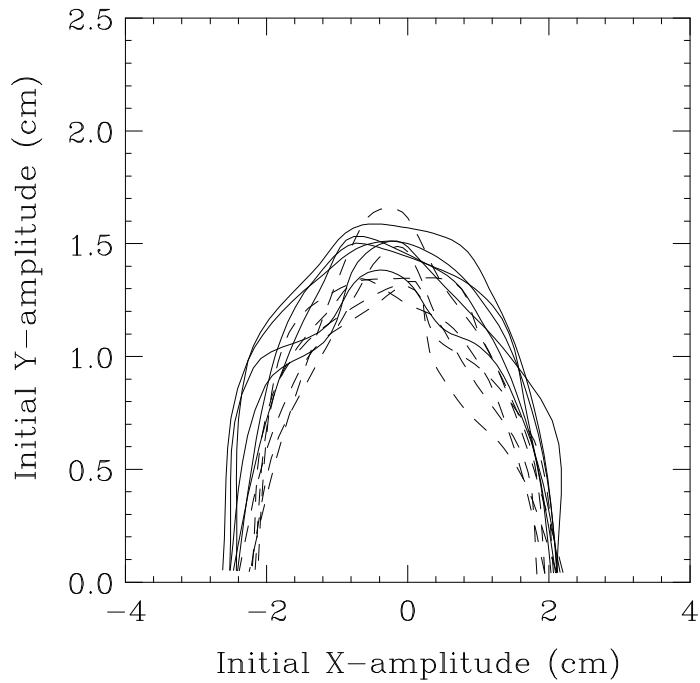}
\vspace{95mm}
\caption{Dynamic Aperture with Two Family Harmonic Sextupoles in
Arc Cells with SPEAR~3 Machine Errors for 0~(solid) and 
3\%~(dash) Momentum Oscillations.}
\label{y:f:hsxseed}
\end{figure}

Based on formulas in~\cite{y:hsex}, we analyzed the magnitude of harmonic 
components in the amplitude dependent tune shift generated by the chromatic
sextupoles. The analysis showed that one of the largest contributions comes 
from 14th and 18th horizontal tune harmonics. To minimize these 
contributions we tested a scheme which has two family harmonic sextupoles
in each of the 14 arc cells. Because of very limited space in non-dispersive
region in the cells, in this test we used thin lens harmonic sextupoles
attached to cell quadrupoles in a doublet. The strength of these sextupoles
was optimized by scanning and maximizing dynamic aperture. The reduction
of tune shift with amplitude due to harmonic sextupoles was verified using 
HARMON and the results are presented in Table~6, where $\epsilon$ is the
rms beam emittance. The dynamic aperture calculation with the harmonic 
sextupoles for error free lattice and machine with errors is shown in 
Fig.~\ref{y:f:hsxbare},\ref{y:f:hsxseed}. Compared to 
Fig.~\ref{y:f:bare}, the improvement of error free dynamic 
aperture is about 10-15\%. However, with machine errors included the 
improvement of horizontal aperture reduces to a minimum. Taking into 
account the cost and design complications associated with additional 
sextupoles as well as marginal aperture improvement, the harmonic 
sextupole correction was not included in the current design.

\begin{table}[htb]
\caption{Amplitude Dependent Tune Shift for SPEAR~3 with and w/o
Two Family Harmonic Sextupoles in Arc Cells.}
\label{y:t:hsex}
\begin{center}
\begin{tabular}{lccc}
\textbf{Amplitude dependent} & \boldmath{$\underline{d\nu_{x}}$}
                             & \boldmath{$\underline{d\nu_{y}}$}
                             & \boldmath{$\underline{d\nu_{y}}$} \\
\textbf{tune shift} & \boldmath{$d\epsilon_x$}
                    & \boldmath{$d\epsilon_y$}
                    & \boldmath{$d\epsilon_x$} \\
\tableline
w/o harmonic sextupoles & 878 & 1392 & 643 \\
with harmonic sextupoles & 446 & 866 & 479
\end{tabular}
\end{center}
\end{table}

\subsection*{Insertion Devices} 
At present, seven horizontally deflecting wigglers are planned
in the 3~m drifts between arc cells. The typical parameters of
these insertion devices are: the peak field $B_y$ up to 2~T, and the total
length up to 2.3~m. Though the wiggler magnetic field is highly non-linear,
some of its high order effects are locally canceled due to wiggler
periodicity. The remaining lowest order perturbations to the beam optics are
vertical focusing and amplitude-dependent vertical tune shift due to 
octupole-like horizontal field. In summary, the wiggler effect on dynamic 
aperture can arise from perturbation of $\beta$ functions and tune, 
high order field effects, reduced periodicity and symmetry of the lattice, 
as well as wiggler field errors and misalignment.

In SPEAR~3 the wiggler perturbation of $\beta$ functions and phase advance 
will be locally corrected using doublet quadrupoles in the cells adjacent 
to either side of the wiggler~\cite{y:wig}. Tracking studies with seven
wigglers showed that without high order wiggler effects, the corrected 
wiggler focusing alone does not reduce dynamic aperture. Without this 
correction the vertical aperture reduces by about 20\%, though it is still 
well outside the $y=\pm6$~mm wiggler physical aperture.

\begin{figure}[tb]
\includegraphics{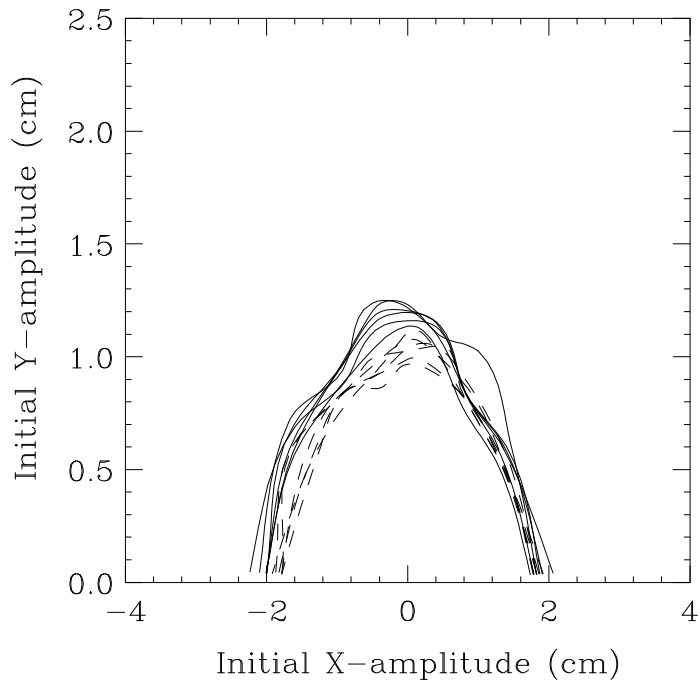}
\vspace{95mm}
\caption{Dynamic Aperture with 7~Wigglers, Corrected Wiggler
Focusing, Systematic Wiggler Multipole Errors and 6 Seeds of Random
Machine Errors for 0~(solid) and 3\%~(dash) Momentum Oscillations.
The Intrinsic High Order Wiggler Fields not Included.}
\label{y:f:wig}
\end{figure}

The study of wiggler multipole errors was based on recently measured
systematic field errors in one of the strongest wigglers (Beamline~11).
The measured field data was fit to a set of normal and skew field multipoles 
up to 12th order and used as wiggler errors in the tracking. 
Fig.~\ref{y:f:wig} shows that when these systematic multipole errors are 
included, the dynamic aperture with seven wigglers reduces by about 10\%. 

The effects of the non-linear wiggler field can further reduce the aperture. 
Since these intrinsic high order fields are in the horizontal plane, they
mostly affect the vertical aperture. Simulation of the first two 
non-linear terms (octupole and dodecapole-like field) showed rather 
modest reduction of vertical aperture from 11~mm to 9~mm with the above 
systematic multipole errors included~\cite{y:wig}. This aperture is still 
well outside the wiggler physical aperture.
\subsection*{Effect of Lattice Periodicity}
\begin{figure}[tb]
\includegraphics{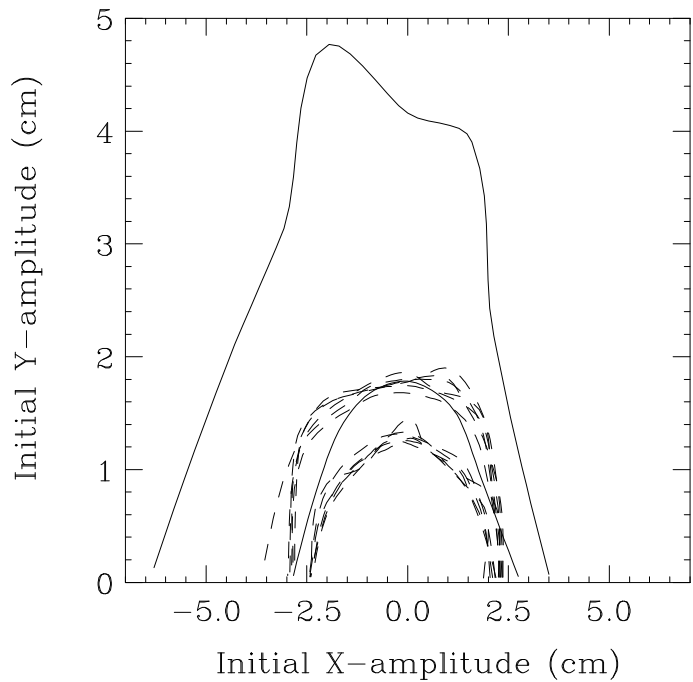}
\vspace{95mm}
\caption{On-Momentum Dynamic Aperture: 1)~14~Cells w/o Errors (bigger solid
curve), 2)~Full Lattice w/o Errors (smaller solid), 3)~14~Cells with
Errors (bigger dash), 4)~Full Lattice with Errors (smaller dash).}
\label{y:f:period}
\end{figure}

The advantage of a 
high periodicity lattice is that the number of resonances 
excited by systematic multipole field errors is reduced in proportion 
to the number of periods. In SPEAR~3 the matching cells break the
14~periodic arc cells and reduce the machine periodicity to~2.
To verify the effect of periodicity on dynamic aperture, we tracked
particles in a lattice which had only 14~identical arc cells and 
compared the results to the aperture of the full lattice. The results 
for on-momentum particles are shown in Fig.~\ref{y:f:period}. 
The two solid curves correspond to
aperture without machine errors, and the dash curves define the aperture
for 6 seeds of machine errors. In both cases the pure 14 periodic cells 
provide a larger dynamic aperture compared to the nominal lattice
with matching cells. Since without errors the only non-linear fields
are from sextupoles, one can conclude that cancellation of systematic 
sextupole aberrations is much better in a more periodic lattice. 
Another observation is that either breaking periodicity or 
including random machine errors will
reduce the SPEAR~3 aperture to about similar size. 

Due to geometric constraints in the existing SPEAR tunnel it was unavoidable
to break the periodicity of arc cells. To keep the effective periodicity
of~14, one of the earliest proposals suggested a $I$-transformation for the
matching cell lattice between the arcs~\cite{y:garren}. However, the
matching cells also contribute a significant amount of chromaticity, and
it was found that relaxing the matching cell optics improved the off-momentum
dynamic aperture without compromising the on-momentum aperture.
\section*{Conclusions}
A long Touschek lifetime and adequate injection conditions were the
primary motivations to maximize the SPEAR~3 dynamic aperture. The
described optimization included linear optics, working tune, 
chromaticity and coupling correction, and compensation of wiggler 
focusing. Other potential improvements were analyzed as well.
The effects of large momentum oscillations, realistic machine errors, 
insertion devices and larger optics perturbations on dynamic aperture 
were verified. The results consistently showed that the dynamic aperture
for up to $\delta=3\%$ momentum errors is not significantly
affected by these effects. The presented analysis shows that SPEAR~3 
dynamic aperture will provide adequate injection conditions and result 
in $>100$~hrs of Touschek beam lifetime.
\section*{Acknowledgements}
The authors would like to thank many people who contributed to the 
SPEAR~3 lattice design. In particular, the authors would like to thank 
M.~Cornacchia, Y.~Cai, A.~Garren, R.~Hettel, J.~Safranek and 
H.~Wiedemann for many fruitful discussions.


\begin{thebibliography}{99}
\bibitem{y:grad} The use of gradient dipoles for SPEAR~3 was proposed by
Safranek,~J.
\end{thebibliography}
\end{document}